# "Big Data" and its Origins

Francis X. Diebold
University of Pennsylvania

The "Big Data" phenomenon, by which I mean the explosive growth in data volume, velocity, and variety, is at the heart of modern science (and is similarly central to modern business). Indeed, the necessity of grappling with Big Data, and the desirability of unlocking the information hidden within, is now a key theme in all the sciences – arguably the key scientific theme of our times.

Parts of my field of econometrics, to take a tiny example, are working furiously to develop methods for learning from the massive amount of tick-by-tick financial market data now available.[1] In response to a question like "How big is your data set?" in a financial econometric context, an answer like "90 observations on each of 10 variables" would have been common fifty years ago, but now it's comically quaint. A modern answer is likely to be a file size rather than an observation count, and it's more likely to be 200 gigabytes (GB) than the 5 kilobytes (say) of 50 years ago. Moreover, someone reading this in 20 years will surely have a good laugh at my implicit assertion that a 200 GB dataset is large. In other disciplines like physics, 200 GB is already small. The Large Hadron Collider experiments that led to discovery of the Higgs boson, for example, produce a petabyte ($10^{15}$ bytes) of data *per second*.

My interest in the origins of the term "Big Data" was piqued in 2012 when Marco Pospiech, at the time a PhD student studying the Big Data phenomenon at the Technical University of Freiberg, informed me in private correspondence that he had traced the use of the term (in the modern sense) to my paper, "'Big Data' Dynamic Factor Models for Macroeconomic Measurement and Forecasting", presented at the Eighth World Congress of the Econometric Society in Seattle in August 2000.[2,3]

Intrigued, I did a bit more digging. And a deeper investigation reveals that the situation is more nuanced than it first appears.

**Big Data and me**

I stumbled on the term "Big Data" innocently enough, via discussion of two papers that were also presented at the World Congress of the Econometric Society in August 2000.[4,5] These papers took a new approach to macro-econometric dynamic factor models (DFMs): simple statistical models in which variation in a potentially large set of serially-correlated observed variables is driven in part by their dependence on a small set of underlying serially-correlated latent variables, or "factors". DFMs are popular in dynamic economic contexts, in which observed variables often move closely together.[6]

Early-vintage DFMs included just a few variables, because parsimony was essential for tractability of numerical likelihood optimization.[7] The new work, in contrast, showed how DFMs could be estimated using statistical principal components in conjunction with least squares regression, thereby dispensing with numerical optimization and opening the field to analysis of much larger data sets while nevertheless retaining a likelihood-based approach.

My discussion had two overarching goals. First, I wanted to contrast the old and new macro-econometric DFM environments. Second, I wanted to emphasize that the driver of the new macro-econometric DFM developments matched the driver of many other recent scientific developments:

explosive growth in available data. To that end, I wanted a concise term that conjured a stark image. I settled on the term "Big Data", which seemed apt and resonant and intriguingly Orwellian (especially when capitalized), and which helped to promote both goals.

**Murky origins**

My paper seems to have been the first academic reference to Big Data in a title or abstract in the statistics, econometrics, or additional *x*-metrics (insert your favorite *x*) literatures. Moving backward from there, things get murkier. It seems academics were aware of the emerging phenomenon but not the term.[8] Conversely, a few pre-2000 references, both academic and non-academic, used the term but were not thoroughly aware of the phenomenon.

On the academic side, for example, Tilly (1984) mentions Big Data, but this article is not about the Big Data phenomenon and demonstrates no awareness of it; rather, it is a discourse on whether statistical data analyses are of value to historians.[9] On the non-academic side, the margin comments of a computer program posted to a newsgroup in 1987 mention a programming technique called "small code, big data" (note the absence of capitalization; bit.ly/3r3sdIJ).

Next, Eric Larson provides an early popular-press mention in a 1989 *Washington Post* article about firms that assemble and sell lists to junk-mailers. He notes in passing that: "The keepers of Big Data" – now capitalized – "say they do it for the consumer's benefit."[10] Finally, a 1996 PR Newswire, Inc. release mentions network technology "for CPU clustering and Big Data applications…".

There is, however, some pre-2000 activity that is spot-on. First, on the industry side, "Big Data" the term, coupled with awareness of "Big Data" the phenomenon, was clearly percolating at Silicon Graphics (SGI) in the mid 1990s. John Mashey, retired former chief scientist at SGI, produced a 1998 SGI slide deck entitled "Big Data and the Next Wave of InfraStress", which clearly demonstrates this awareness (bit.ly/34lGqWS). Related, SGI ran magazine ads that featured the term "Big Data" in *Black Enterprise* in 1996, several times in *Info World* starting November 1997, and several times in *CIO* starting February 1998. Clearly, Mashey and the SGI community were on to Big Data early, using it both as a unifying theme for technical seminars and as an advertising hook.

Second, on the academic side, in the context of computer graphics, Cox and Ellsworth (1997) describe "an interesting challenge for computer systems: data sets are generally quite large, taxing the capacities of main memory, local disk, and even remote disk", which they call "the problem of big data".[11] In addition, Weiss and Indurkhya (1998) note that "… very large collections of data … are now being compiled into centralized data warehouses, allowing analysts to make use of powerful methods to examine data more comprehensively. In theory, 'Big Data' can lead to much stronger conclusions for data-mining applications, but in practice many difficulties arise."[12]

Finally, arriving on the scene later but also going beyond previous work in compelling ways, Laney (2001) highlighted the "Three V's" of Big Data (Volume, Variety and Velocity) in an unpublished 2001 research note at META Group.[13] Laney's note is clearly relevant, producing a substantially enriched conceptualization of the Big Data phenomenon. In short, if Laney arrived slightly late, he nevertheless brought more to the table.

The rest, as they say, is history. As the synopsis for a recent BBC Radio 4 broadcast (bbc.in/3agQcgQ) puts it:

> In 2012, Big Data entered the mainstream when it was discussed at the World Economic Forum in Davos. In March that year, the American government provided $200 million in

research programs for Big Data computing. Soon afterward, the term was included in the Oxford English Dictionary for the first time.

**A discipline, and a triumph?**

Big Data is arguably now not only a phenomenon and a well-known term, but also a discipline. To some, that might sound like marketing fluff, and it can be hard to resist smirking when told that major firms are rushing to create new executive titles like "Vice President for Big Data" (nyti.ms/3oZ1lXL). But the phenomenon behind the term is real, so it may be natural for a corresponding new business discipline to emerge, whatever its executive titles.

Business discipline or not, it's still not obvious that Big Data constitutes a new *scientific* discipline. Sceptics will argue that traditional areas like statistics and computer science are perfectly capable of confronting the new phenomenon, so that Big Data is not a new discipline, but rather just a box drawn around some traditional ones. It's hard not to notice, however, that the whole of Big Data seems greater than the sum of its parts. That is, by drawing on perspectives from a variety of traditional disciplines, Big Data is not merely taking us to (bigger) traditional places; rather, it's taking us to very new places, unimaginable only a short time ago. In a landscape littered with failed attempts at interdisciplinary collaboration, Big Data is an interdisciplinary triumph.

As always, however, there's a flip side. Big Data pitfalls may lurk, for example, in the emergence of continuous surveillance facilitated by advances in real-time massive data capture, storage, and analysis. As George Orwell wrote in his famously prescient novel, *Nineteen Eighty-Four*:

> Always eyes watching you and the voice enveloping you. Asleep or awake, indoors or outdoors, in the bath or bed – no escape. Nothing was your own except the few cubic centimeters in your skull.[14]

Time will reveal how Big Data opportunities and pitfalls evolve and interact.

**About the author**

Francis X. Diebold is the Paul F. Miller, Jr. and E. Warren Shafer Miller professor of social sciences, and professor of economics, finance and statistics at the University of Pennsylvania.

**Acknowledgments**
For helpful comments, I thank (without implicating anyone in any way): Larry Brown, David Cannadine, Xu Cheng, Tom Coupe, Flavio Cunha, Susan Diebold, Melissa Fitzgerald, Dean Foster, Michael Halperin, Steve Lohr, John Mashey, Tom Nickolas, Lauris Olson, Mallesh Pai, Marco Pospiech, Brian Tarran, Frank Schorfheide, Minchul Shin, Mike Steele and Stephen Stigler.